\documentclass[aps,prl,preprint,groupedaddress]{revtex4}
\usepackage[pdftex]{graphicx}

\begin{document}

\title{Phase Transition in Sexual Reproduction and Biological Evolution}

\author{Marta Zawierta}
\author{Wojciech Waga}
\author{Dorota Mackiewicz}
\author{Przemys{\l}aw Biecek}
\author{Stanis{\l}aw Cebrat}
\email{cebrat@smorfland.uni.wroc.pl}

\affiliation{Department of Genomics, Faculty of Biotechnology, Wroc{\l}aw University, ul. Przybyszewskiego 63/77, 51-148 Wroc{\l}aw}

\date{\today}

\begin{abstract}
Using Monte Carlo model of biological evolution we have discovered that populations can switch between two different strategies of their genomes' evolution; Darwinian purifying selection and complementing the haplotypes. The first one is exploited in the large panmictic populations while the second one in the small highly inbred populations. The choice depends on the crossover frequency. There is a power law relation between the critical value of crossover frequency and the size of panmictic population. Under the constant inbreeding this critical value of crossover does not depend on the population size and has a character of phase transition. Close to this value sympatric speciation is observed. \end{abstract}


\maketitle

\section{Introduction}
According to neo-Darwinism, biological evolution is driven by two random processes - mutations and recombinations and one directional - selection \cite{Ayala}. While in general it is true, the reality is much more complicated - mutations and recombinations are highly biased in both, frequencies and locations while selection, in many instances, does not see the defective alleles in the diploid genomes. In this paper we are going to describe the phenomena associated with the interplay between the intragenomic recombination rate and inbreeding coefficient. Inbreeding is a measure of genetic relation between individuals in the population. In the highly inbred populations the probability of meeting two relatives as sexual partners for reproduction is relatively high. If we look at the very simple genealogical tree (Fig. 1) we can see that in the second generation it is possible that two identical haplotypes can meet in one genome and if there are some defective alleles in these haplotypes (a,b,c,d), they can appear in both homologous loci in the diploid genome determining the deleterious phenotype (Fig. 1, F2, example 2). The probability of such situation is much lower in the large panmictic populations where it is very unlikely to choose the close relative individual as a partner for reproduction (in the panmictic populations partners for reproduction are randomly assorted from the whole population). In Fig. 1 we have shown two examples - one when two haplotypes which already coexisted in one genome in the past meet in the offspring (F2, example 1) and the other one when two identical haplotypes meet in the offspring genome (F2, example 2). The first situation is more advantageous because this configuration has been already tested by selection in the earlier generation and all defcts are complemented by the wild alleles (A,B,C,D). The second one is more risky because homozygous defective loci appear if there are some defective alleles in the haplotypes. Thus, for choosing the proper haplotype in the highly inbred populations the haplotype recognition during the fertilization would be an advantageous strategy \cite{Stauffer}. On the other hand, recombination between haplotypes during the gamete production (meiosis) disturbs this process and even gamete recognition would not help a lot under higher recombination rate (Fig. 1 example 3). In fact, populations can evolve under two different conditions: 
\begin{itemize}
\item 	in a case of the very small population with low intragenomic recombination rate the more advantageous strategy is to form the genomes of the complementing haplotypes, in such genomes more defective recessive alleles can be compensated or complemented by the wild alleles,
\item 	in the larger populations, the probability of inheriting from both parents the complementing haplotypes originated from the common ancestor is low and the defective alleles are eliminated by purifying Darwinian selection, rather \cite{Zawierta}.
\end{itemize}

In the Nature the size of population does not directly determine the inbreeding because individuals do not look for a partner in the whole large population but in the much more restricted vicinity - populations are not panmictic. To study this effect, simulations of the populations' evolutions were performed on the square lattice with the declared maximum distances between sexual partners and between parents and the offspring. At the edges of such expanding populations the inbreeding is higher and complementation strategy prevails while in the central parts inbreeding is lower and purifying Darwinian selection prevails. Such conditions favour fast sympatric speciation \cite{Waga}. Sympatric speciation is an emergence of a new species inside the old one without any physical, geographical or ecological barriers. Since Mayr \cite{Mayr} who was rather sceptical about the possibility of speciation in sympatry, the probability of such a speciation is still a very hot and debatable problem in the evolutionary biology \cite{Jiggins}. Nevertheless, some biological evidences for sympatric speciation are already known \cite{Barluenga}, \cite{Bunje}. Also computer modelling has shown the feasibility of speciation in sympatry \cite{Doebeli}, \cite{Luz}, \cite{Stauffer2}, \cite{Waga}. It has been shown in the computer studies that small populations evolving independently under relatively low recombination rate behave as different species and characterize with outbreeding depression - no survival or decreased survival of hybrids formed by crossing individuals from two independent populations \cite {Aga}, \cite{Bonkowska}. Such an effect was not observed for large populations with high recombination rate. On the other hand, the results obtained in the simulations on lattice suggested that the interplay between the intragenomic recombination rate and inbreeding (not the total population size) has a deciding role in the process of speciation.

To show the relations between the population size, inbreeding coefficient and intragenomic recombination rate in the process of speciation we have simulated the evolution of panmictic populations and population on the square lattice with the defined inbreeding under different intragenomic recombination rate. 

\subsection{Model of evolution of panmictic populations}

Populations are composed of individuals represented by their diploid genomes. Each genome consists of two haplotypes (bitstrings), 64 bits long. Bits set to 0 correspond to the wild (correct) alleles while bits set to 1 correspond to defective alleles. All defective alleles are recessive which means that both alleles at the corresponding positions (the same locus) have to be defective to determine the deleterious phenotype. To reproduce, two randomly chosen individuals replicate their haplotypes and introduce into each copy a new mutation in the randomly chosen locus. Mutation replaces 0 by 1, if a bit in the chosen locus is 1 it stays 1 (there are no reversions). The mutated copies of haplotypes recombine in the randomly chosen point with the declared probability (process mimicking crossover in meiosis). The end products of these processes correspond to gametes. A gamete of one individual is joined with a gamete of the other one to form the offspring. A newborn dies because of genetic death if at least one locus determines the defective phenotypic character (it corresponds to a zygotic death - the individual has no chance to produce any offspring). If the offspring has no defective phenotypic traits it survives. At the beginning of each Monte Carlo step 2\% of individuals are killed randomly. This 2\% gap is filled up with the new offspring. The logistic Verhulst equation to control the population size was not used because the population size intentionally was kept constant. Populations start with perfect genomes - all bits set to 0. Size of populations ranged from 100 to 10,000 individuals.

\subsection{Model of population evolution on a square lattice }

In this version the populations are composed of male and female individuals, each represented by two haplotypes. One square of lattice can be occupied at most by one individual. A female, to reproduce, has to find a partner at distance $P\leq 2$ and she places a newborn at the distance $B\leq 2$ if there is a free place for it; otherwise an offspring is not born. At each Monte Carlo step 2\% of population is randomly killed to make a room for newborns. To avoid the border conditions, the square lattice has been wrapped on the torus. Size of lattices ranged from $128 \times 128$ to $1024 \times 1024$. All other parameters are like in the panmictic model.

\section{Results}
\subsection{Panmictic populations}
In the first series of simulations we have checked how the fraction of defective genes in the genomes depends on the population size and the crossover rate. The results are shown in Fig. 2. The crossover rate where the fraction of defective alleles drastically changes ("transition point") is higher for smaller populations than for larger ones. In Fig. 3 the values of recombination frequencies for the "transition points" were plotted against the sizes of populations.  Results show the power law relations between the transition point values and the sizes of populations. For low recombination rates, the fractions of defective genes in the genomes reach 0.5. If the distribution of defects along the haplotypes is random, the probability of forming the survival offspring would be negligible and population should die. To escape the extinction the only possibility is to complement the defective alleles. The whole haplotypes, or at least long stretches of bits, should be complementary. However, recombination between haplotypes disturbs such a complementation. Thus, the results suggest the prominent role of intragenomic recombination rate (crossover) in the establishment of the complementing strategy. Populations evolving under combination of parameters above the curve in Fig. 3 are under the purifying Darwinian selection. For combination of parameters below the curve the complementing strategy is more advantageous. To study the phenomena close to transition point the reproduction potential of populations in this region have been examined. To measure the reproduction potential, the average number of births per one surviving offspring has been counted (Fig. 4). One can notice that at the critical value of recombination this survival probability is the lowest. Thus, the selection should favour the recombination rate higher or lower than the critical value. At the higher recombination rate the Darwinian purifying selection operates but at the lower recombination - the complementation strategy is prevails. 

All results presented in Figs. 2-4 were obtained in simulations of panmictic populations. They suggest that the recombination frequency where the population switches between the two possible strategies depends on the population size. In fact in such a kind of simulations the size of population determines the inbreeding coefficient. In the Nature, individual, instead of looking for sexual partner in the whole populations, chooses a partner in the close neighbourhood, rather. Thus, the effective size of population in respect to the random mating is much smaller. To introduce this into the model we have performed all further simulations on the square lattice.

\subsection{Evolution on the square lattice}

To show that the value of recombination rate at the transition point does not depend on the population size but on the inbreeding, rather a series of simulations on square lattice were performed (see model section for details). In all simulations females looked for partners at distance $P\leq 2$ and placed a newborn at the distance $B \leq 2$. These parameters stabilized the inbreeding during the simulations. The transition point for the inbreeding described by these two parameters is close to 0.25 independently of the population size and the plots describing the fraction of defective alleles in three different sizes of populations evolving under different recombination rate overlap (see Fig. 5). Since the transition does not depend on population size it could be called the phase transition. 

\subsection{Characteristics of the two phases}

The first difference between the two phases - for recombinations below and above the transition point is the fraction of defective genes in the genomes of populations. For lower recombination rate it is high, suggesting the complementation effect. To show the effect of complementation, the distribution of Hamming distances between two haplotypes in the genomes was prepared. In this case the Hamming distance shows the number of heterozygous loci (heterozygous means that in a pair of corresponding alleles one is defective and one is wild). Since even one deleterious phenotypic trait is lethal (kills the individual) all loci which are not heterozygous have both alleles correct (wild). The results in Fig. 6 show the distribution of Hamming distances for populations simulated on the torus $256 \times 256$. The maximum of distribution was 49 below the transition point and 24 above the transition point. Note the very narrow range of recombination rate where the shift of the distributions was observed (0.250 - 0.259).

The other important problem, from the biological point of view, is the distribution of defective alleles along the haplotypes. It has been shown in the previous studies that the recombination events accepted by selection are unevenly distributed along the chromosomes - higher probability of acceptance at the subtelomeric regions and lower probability in the central parts \cite{Zawierta} (chromosomes correspond to haplotype in the model). This prediction of the model is supported by the results of the genetic data analyses \cite{Kong}. Such a distribution of recombinations suggests that the ends of chromosomes stay longer under the purifying Darwinian selection than the central parts which mean that different parts of chromosomes can be under different selection pressure. Results of the analyses of the simulation products are shown in Fig. 7. In fact, the distribution of defective alleles (heterozygous loci) is symmetrical; the central parts use the complementing strategy under higher recombination rate even when the subtelomeric parts are already under purifying strategy. Since the complementing strategy could produce differences in the distribution of defective alleles along the haplotypes, these parts should be responsible for speciation. To visualize the distribution of individuals with different genomes we have used the 24 central bits of haplotypes to colour the squares where the corresponding individual are placed. Fig. 8 shows the effect of using the trick of colouring for simulations on the torus $256 \times 256$,$512 \times 512$, $1024 \times 1024$ with $P=B=2$ and recombination rates 0.1. The large spots on the torus after simulation show territories occupied by individuals belonging to the same species. There are no physical borders between them. The demarcation lines are formed because the hybrids generated by parents belonging to the neighbouring species cannot survive. Larger crossover rate or significantly larger distances where sexual partners can be found or the offspring can be placed change the conditions of population evolution and the whole population can behave as a panmictic one \cite{Waga}. More examples in colour can be seen at our web page http://www.smorfland.uni.wroc.pl/sympatry.

\section{Conclusions}

Depending on the internal (crossover rate) and external (inbreeding) conditions  genomes can evolve in two different ways. Large populations with high crossover rate use Darwinian purifying selection. Under such conditions, populations are polymorphic and stable. Small or highly inbred populations with low crossover rate use complementation strategy which could lead to sympatric speciation. Switching between these two strategies has a character of phase transition.

\begin{figure}
\includegraphics[width=0.7\textwidth]{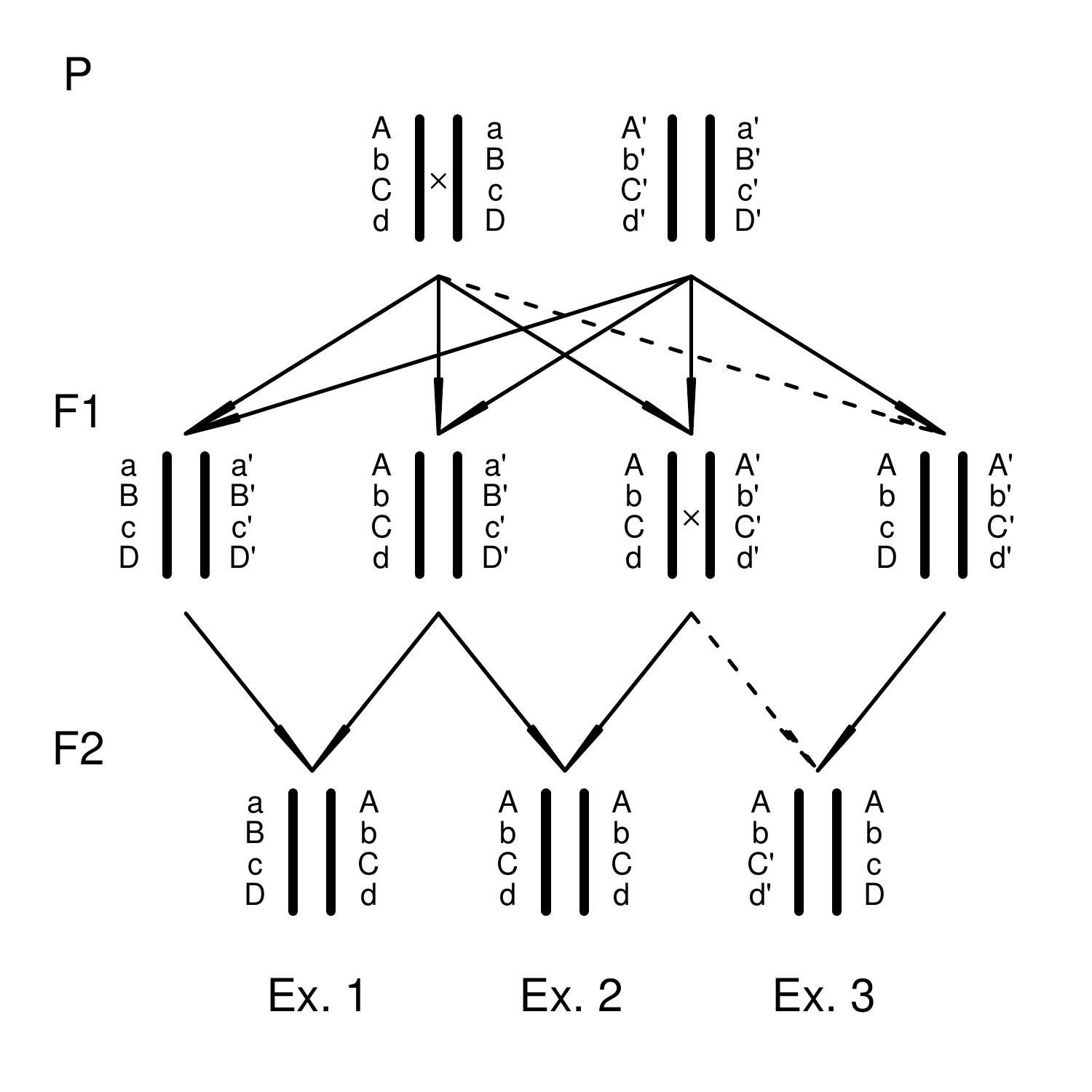}
\caption{\label{fig:figre1}Scheme of the haplotype inheritance during the reproduction of diploids. Assuming that there is no recombination between haplotypes, parents (P) can produce four genetically different offspring (F1). If there is no crossover in the F1 genomes, there is a possibility to reproduce the genome of ancestor in F2 (complementing alleles in the example (1)) or to form the genome of identical haplotypes (example (2) no complementation in two loci). Recombination in F1 (dotted cross) eliminates the possibility of reproducing the genome of complementing haplotypes. The complementation, like in example 1 can be exploited in the highly inbred populations (see text for more details).}
\end{figure}

\begin{figure}
\includegraphics[width=0.7\textwidth]{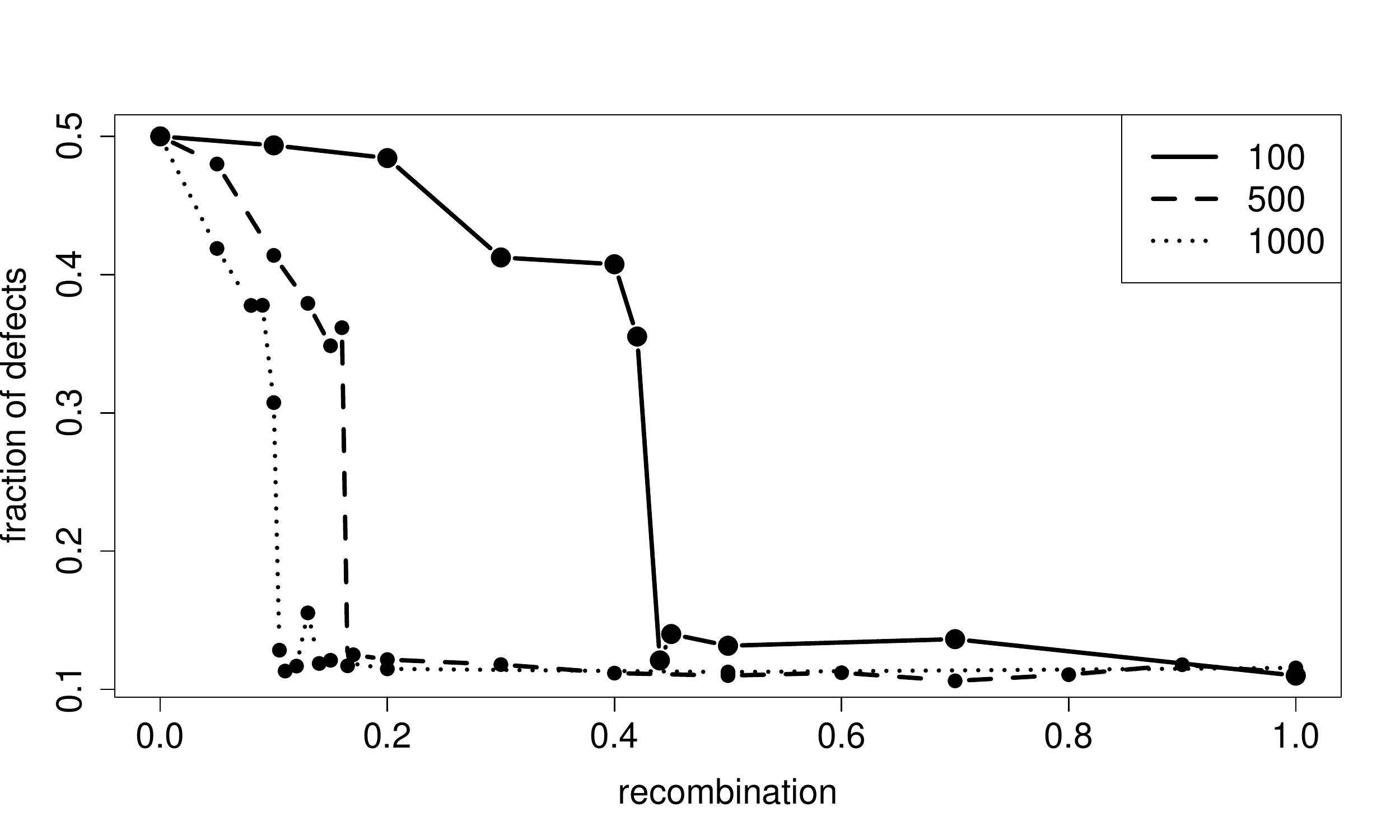}
\caption{\label{fig:figre2} The relation between the fraction of defective genes in the genetic pool and the crossover rate for three different sizes of the panmictic populations: 100, 500 and 1000 individuals (from right to left).}
\end{figure}

\begin{figure}
\includegraphics[width=0.7\textwidth]{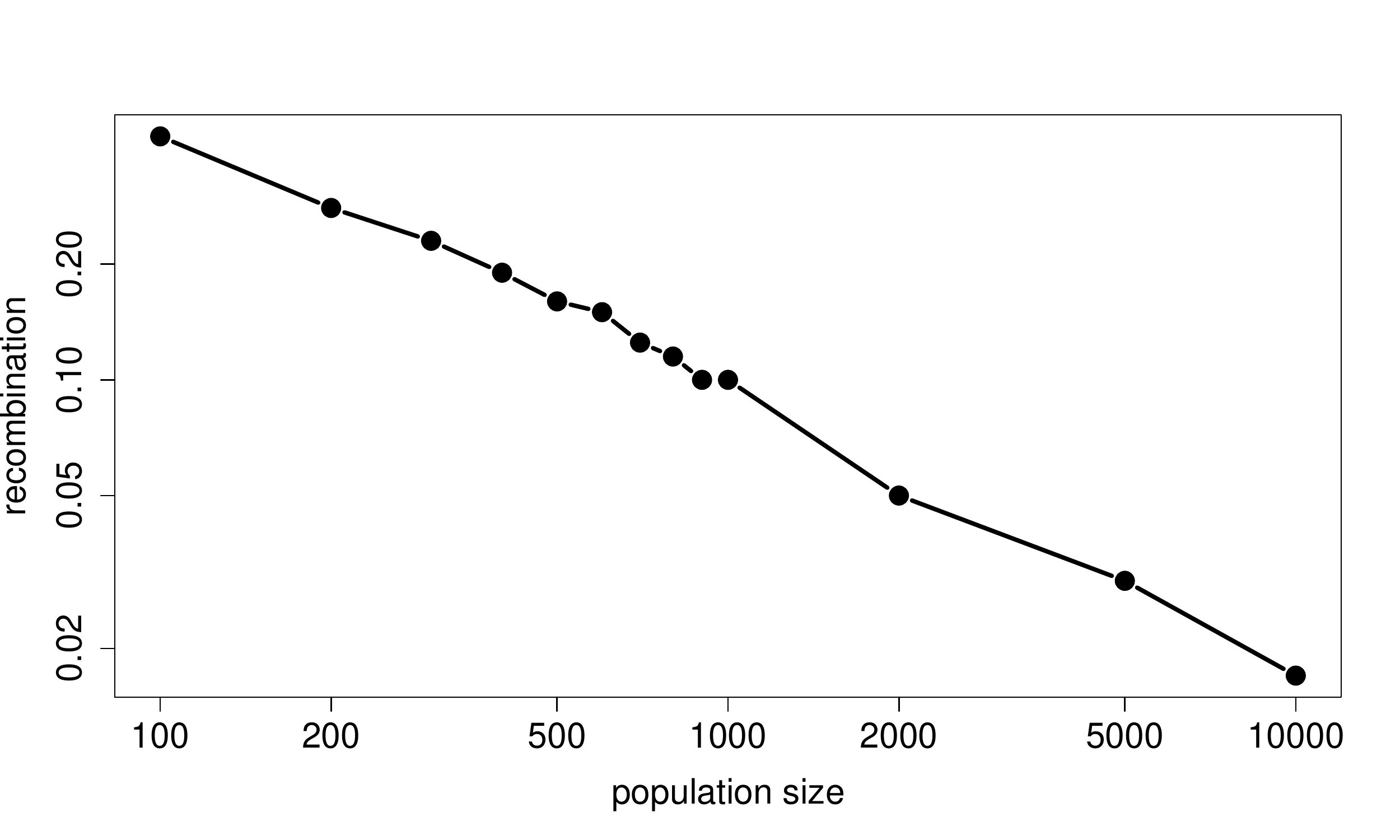}
\caption{\label{fig:figre3} The relation between recombination frequency at the transition point and the population size. Note that data have been plotted in the log/log scale.}
\end{figure}

\begin{figure}
\includegraphics[width=0.7\textwidth]{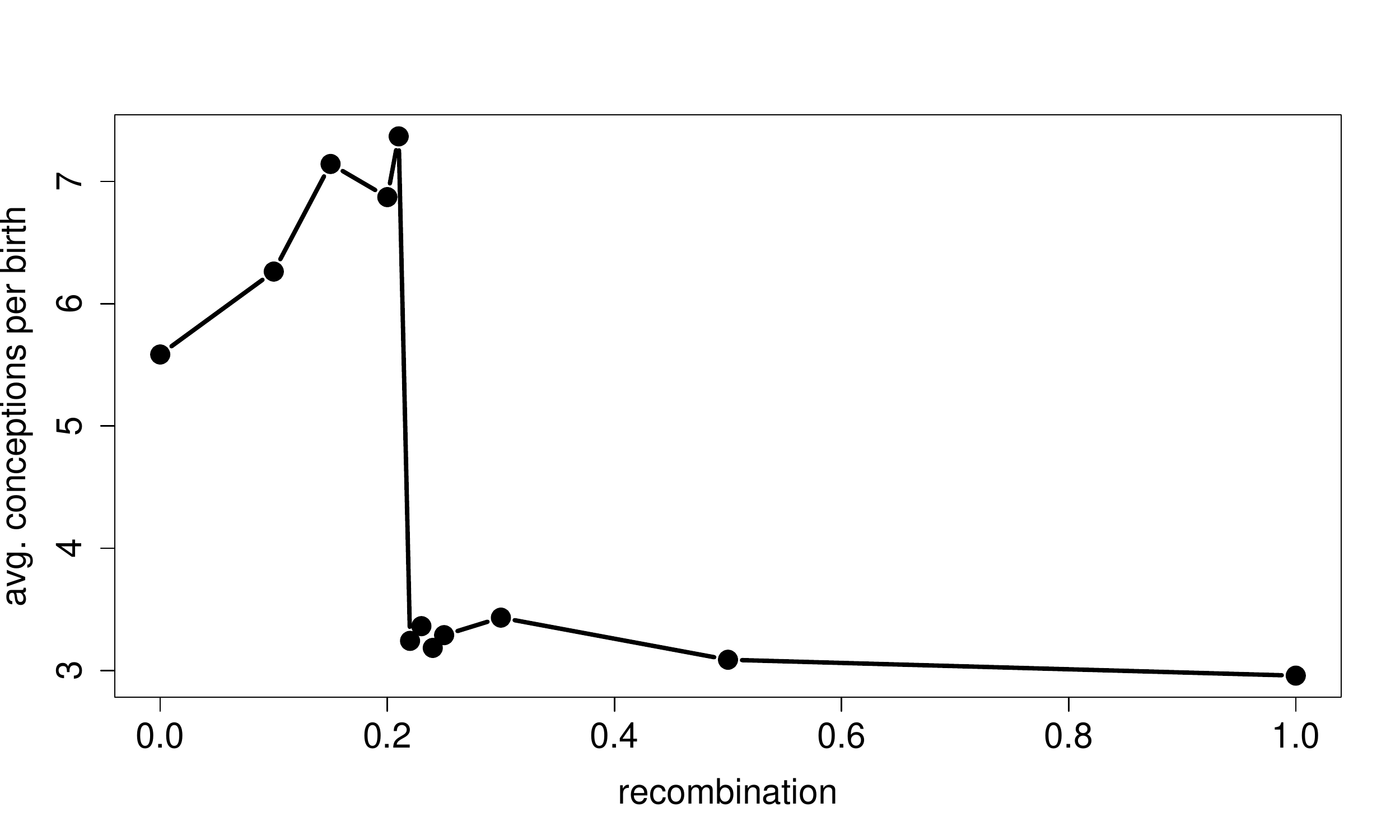}
\caption{\label{fig:figre4} The fertility of populations. Y-axis represent the average number of offspring which has to be produced for one survival; population size - 300 individuals.}
\end{figure}

\begin{figure}
\includegraphics[width=0.7\textwidth]{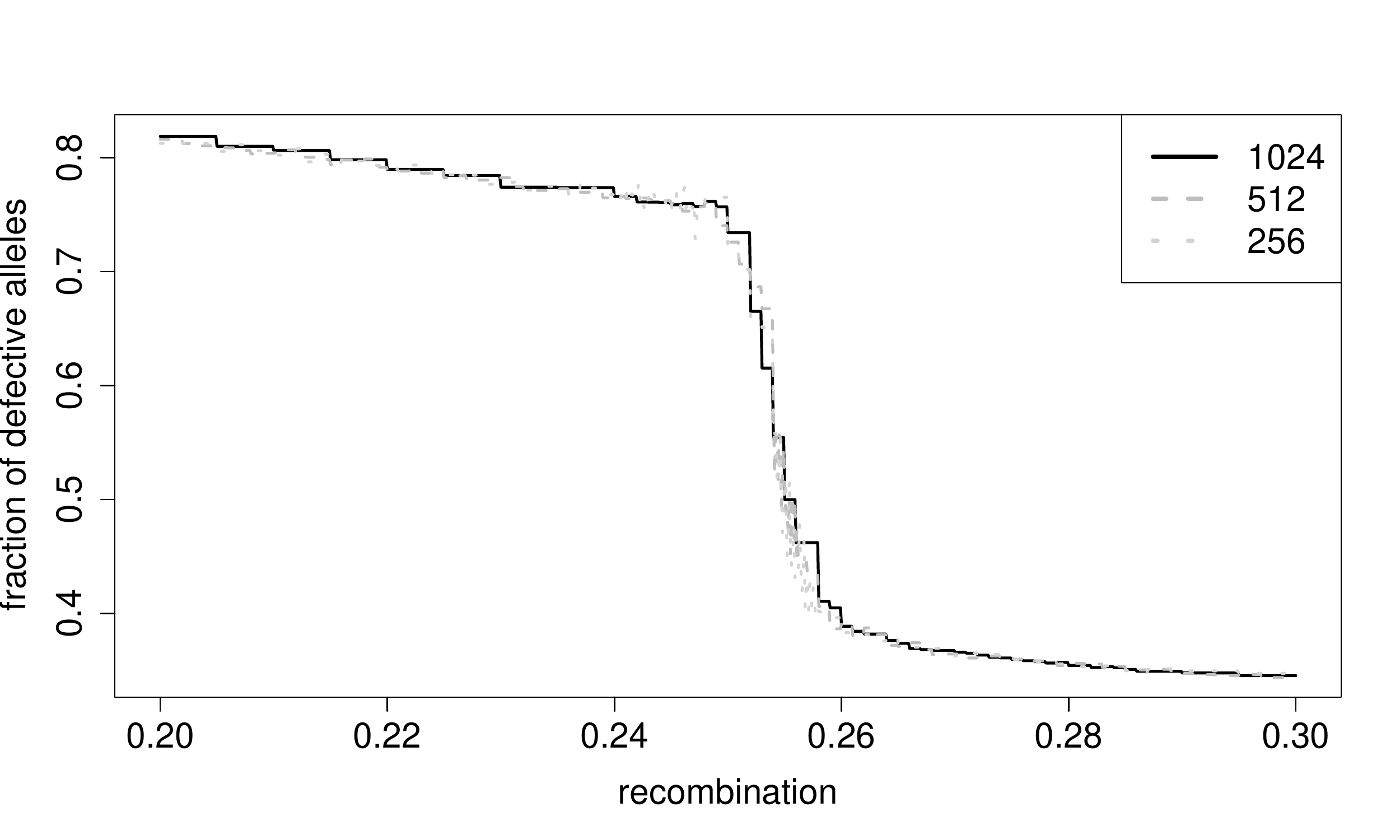}
\caption{\label{fig:figre5} Transition point for different population size with the constant inbreeding coefficient. Simulations were performed on square lattices wrapped on torus (256x256, 512x512, 1024x1024). Inbreeding determined by P=B=2 (see text for details). }
\end{figure}

\begin{figure}
\includegraphics[width=0.7\textwidth]{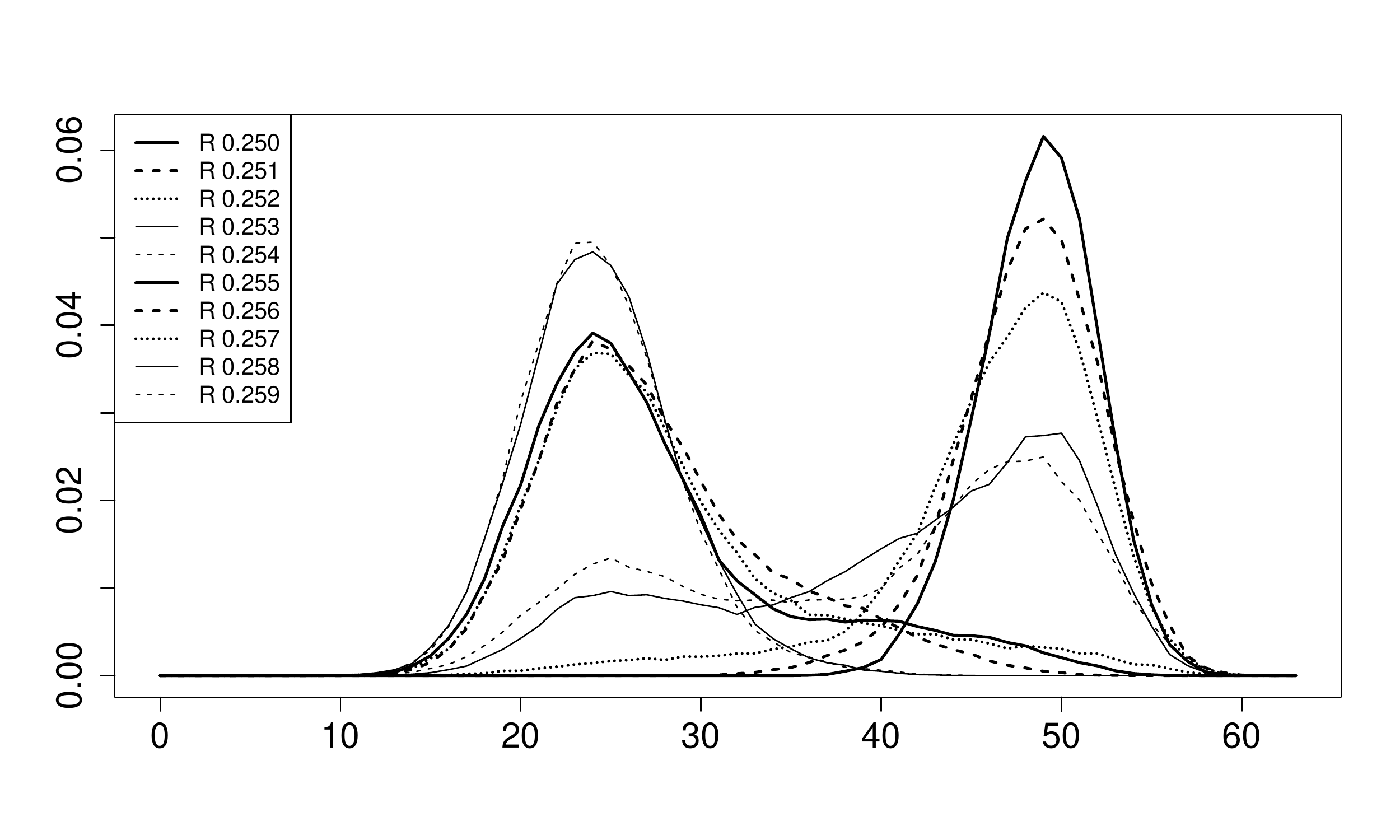}
\caption{\label{fig:figre6} The distribution of Hamming distances for populations simulated on the torus 256x256 under different recombination rate. The maximum of distribution was 49 below the transition point and 24 above the transition point. Note a very narrow range of recombination rate where the shift of the distributions was observed (0.250 - 0.259 from the top to the bottom of the right pick and from the bottom to the top in the left pick). }
\end{figure}

\begin{figure}
\includegraphics[width=0.7\textwidth]{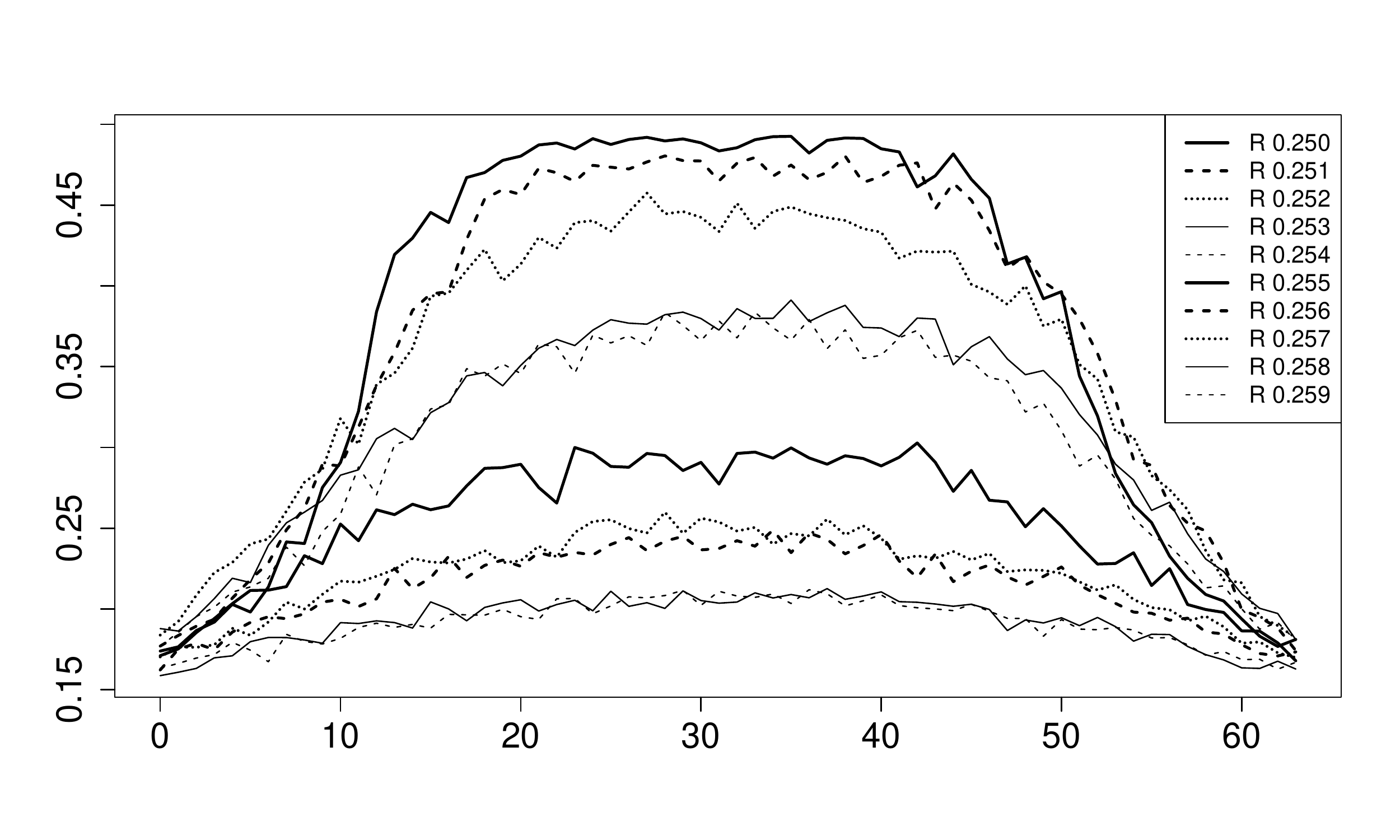}
\caption{\label{fig:figre7} Distribution of defective alleles along the haplotypes; x-axis - the position of locus on haplotype, y-axis - the fraction of defective alleles. Lines describe the distributions for different recombination rates - from 0.250 at the top to 0.259 at the bottom. }
\end{figure}

\begin{figure}
  \begin{picture}(300,300)(10,20)
          \put(0,0){\fbox{\includegraphics[width=300pt]{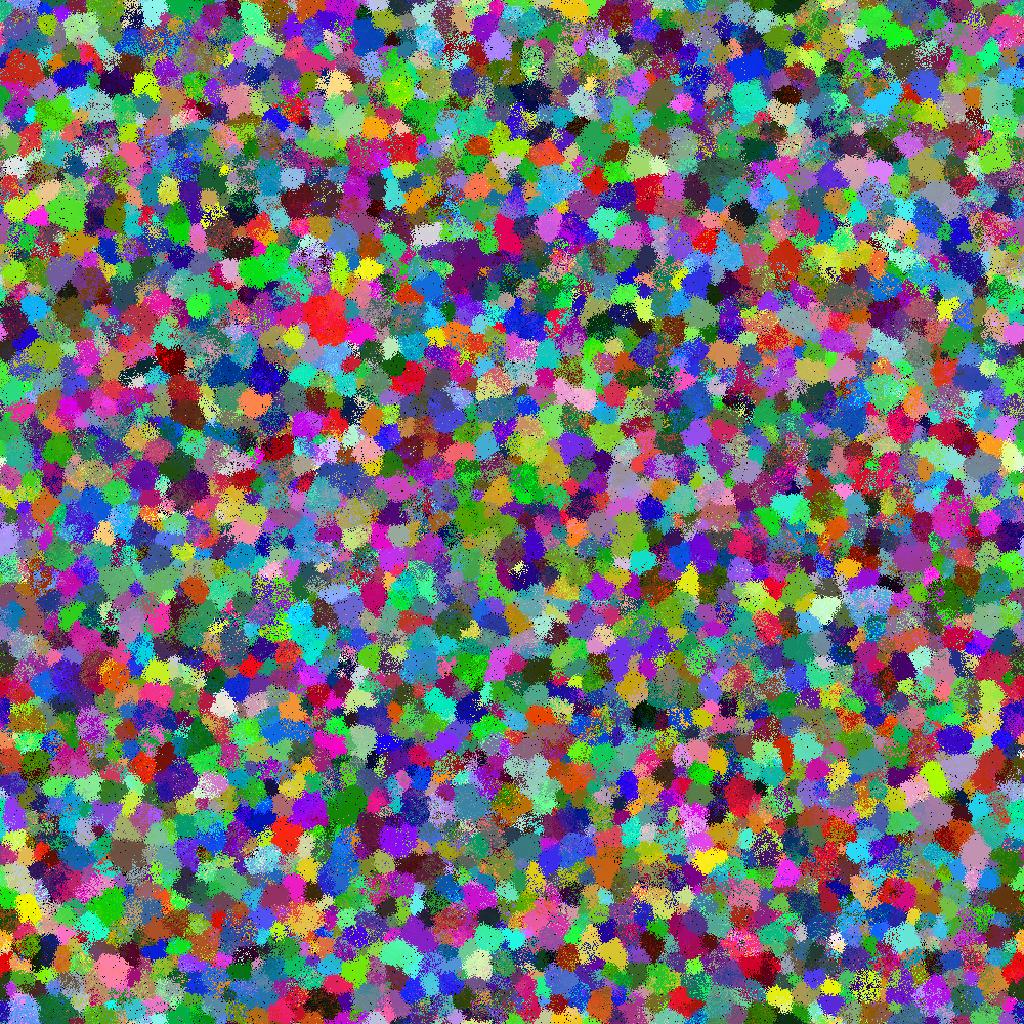}}}
          \put(0,0){\fbox{\includegraphics[width=150pt]{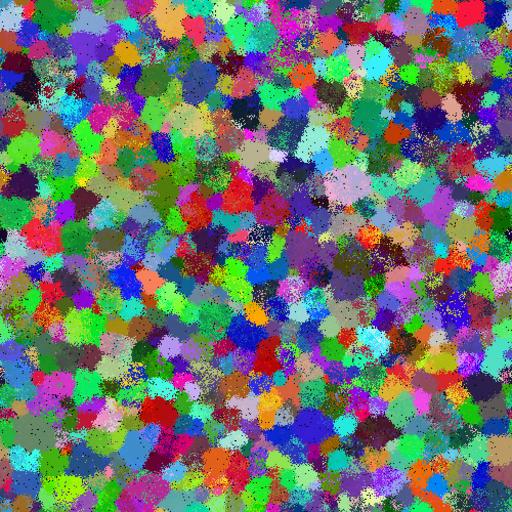}}}
          \put(0,0){\fbox{\includegraphics[width=75pt]{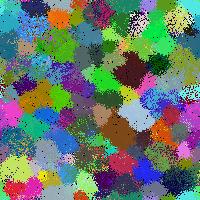}}}
  \end{picture}
 
\caption{\label{fig:figre8} Visualization of speciation effect after simulation of the population evolution on the toruses $256 \times 256$ (small, left bottom),$512 \times 512$(left bottom quarter), $1024 \times 1024$ with $P=B=2$ and recombination rates 0.1. Spots of the same darkness correspond to individuals with the same configuration (distribution) of defective alleles along the central parts of their haplotypes corresponding to one species. At the borders between two species the hybrids are formed but they have no chance to survive. For color pictures and more examples see http://www.smorfland.uni.wroc.pl/sympatry.}
\end{figure}


\begin{thebibliography}{99}
\footnotesize
\bibitem{Ayala} F.J. Ayala and J.A. Kiger, Modern Genetics, The Benjamin/Cumings Pub. Comp. Inc., California (1980).

\bibitem{Stauffer} S. Cebrat and D. Stauffer, Gamete recognition and complementary haplotypes in sexual Penna ageing model, J. Mod. Phys. C. Vol. (in press) arXiv:0709.2420 (2007).

\bibitem{Zawierta} M. Zawierta, P. Biecek, W. Waga, and S. Cebrat, The role of intragenomic recombination rate in the evolution of population's genetic pool, Theory BioSci. 125, 123-132 (2007).

\bibitem{Waga} W. Waga, D. Mackiewicz, M. Zawierta, S. Cebrat, Sympatric speciation as intrinsic property of expanding populations, Theory in Biosciences, 126, 53 - 59 (2007).

\bibitem{Mayr} E. Mayr, Systematics and the Origin of Species. Columbia Press, New York (1942).

\bibitem{Jiggins} C.D. Jiggins, Sympatric speciation: why the controversy? Curr. Biol. 16, R333-R334 (2006).

\bibitem{Barluenga} M. Barluenga, K.N. Stolting, W. Salzbulger, M. Muschick and A. Meyer, Sympatric speciation in Nicaraguan crater lake cichlid fish, Nature 439, 719-723 (2006).

\bibitem{Bunje} P.M.E. Bunje, M. Barluenga and A. Meyer, Sampling genetic diversity in the sympatrically and allopatrically speciating Midas cichlid species complex over a 16 year time series. BMC Evol. Biol., 7, 25-39 (2007).

\bibitem{Aga} A. {\L}aszkiewicz, Sz. Szymczak, Sz. and S. Cebrat, Speciation effect in the Penna aging model, Int. J. Mod. Phys. C 14, 765-774 (2003).

\bibitem{Bonkowska} K. Bonkowska, M. Kula, S. Cebrat and D. Stauffer, Inbreeding and outbreeding depressions in the Penna model as a result of crossover frequency, Int. J. Mod. Phys. C. 18, 1329-1338 (2007). 


\bibitem{Doebeli} M. Doebeli and U. Dieckmann, Speciation along environmental gradients, Nature 421, 259-264 (2003).

\bibitem{Luz} K. Luz-Burgoa, S. Moss de Oliveira, V. Schwämmle and J.S. Sá Martins, Thermodynamic behavior of a phase transition in a model for sympatric speciation, Phys. Rev. E 74, 021910 (2006). 

\bibitem{Stauffer2} D. Stauffer, S. Moss de Oliveira, P.M.C. de Oliveira and J.S. Sá Martins, Biology, Sociology, Geology by Computational Physicists. Amsterdam: Elsevier (2006).

\bibitem{Kong} A. Kong, D.F. Gudbjartsson, J. Sainz, G.M. Jonsdottir, S.A. Gudjonsson, B. Richardsson, S. Sigurdardottir, J. Barnard, B. Hallbeck, G. Masson, A. Shlien, S.T. Palsson, M.L. Frigge, T.E. Thorgeirsson, J.R. Gulcher and K. Stefansson, A high-resolution recombination map of the human genome, Nat. Genet. 31, 241-247 (2002).

\end{thebibliography}
\end{document}